\documentclass[prd,twocolumn,showpacs,amsmath,amssymb,superscriptaddress,preprintnumbers,nofootinbib,showkeys]{revtex4}

\begin{document}

\newcommand{\nablab}{{\mathop {\rule{0pt}{0pt}{\nabla}}\limits^{\bot}}\rule{0pt}{0pt}}

\title{Einstein-aether theory: Dynamics of relativistic particles \\ with spin or polarization in a G\"odel-type universe}

\author{Alexander B. Balakin and Vladimir A. Popov}
 \affiliation{Department of
General Relativity and Gravitation, Institute of Physics, Kazan
Federal University, Kremlevskaya str. 18, Kazan 420008,
Russia}

\date{\today}

\begin{abstract}
In the framework of the Einstein-aether theory we consider a cosmological model, which describes the evolution of the unit dynamic vector field with activated rotational degree of freedom. We discuss exact solutions of the Einstein-aether theory, for which the space-time is of the G\"odel-type, the velocity four-vector of the aether motion is characterized by a non-vanishing vorticity, thus the rotational vectorial modes can be associated with the source of the universe rotation.
The main goal of our paper is to study the motion of test relativistic particles with vectorial internal degree of freedom (spin or polarization), which is coupled to the unit dynamic vector field. The particles are considered as the test ones in the given space-time background of the G\"odel-type; the spin (polarization) coupling to the unit dynamic vector field is modeled using exact solutions of three types. The first exact solution describes the aether with arbitrary Jacobson's coupling constants; the second one relates to the case, when the Jacobson's constant responsible for the vorticity is vanishing; the third exact solution is obtained using three constraints for the coupling constants. The analysis of the exact expressions, which are obtained for the particle momentum and for the spin (polarization) four-vector components, shows that the interaction of the spin (polarization) with the unit vector field induces a rotation, which is additional to the geodesic precession of the spin (polarization) associated with the universe rotation as a whole.

\end{abstract}
\pacs{04.20.-q, 04.40.-b, 04.40.Nr, 04.50.Kd}
\keywords{Alternative theories of gravity, Einstein-aether theory,
unit vector field, universe rotation, spin precession, polarization rotation}
\maketitle

\section{Introduction}

The Einstein-aether theory is a version of the modified gravity based on the introduction of a dynamic vector field $U^i$, which is time-like and unit, i.e., $g_{ik}U^iU^k=1$. This unit vector field realizes the idea of a preferred frame of reference (see, e.g., \cite{N1,N2,N3,CW}), and it characterizes the velocity of the aether motion (see, e.g., \cite{J1,J2,J3,J4,J5,J6,J7,J8,J9,J10} for details, references and analysis of models with various space-time symmetries).
As it was mentioned in \cite{J10}, the integral curves associated with this unit time-like field can be interpreted in terms of a flow of an {\it aether fluid}, the velocity four-vector being tangent to this flow. This aether fluid can be considered in the context of a cosmic {\it dark fluid}. For instance, in the works \cite{Z1,Z2,Z3,EAS1,EAS2} the authors interpret the vector field effects in cosmology as effects of the dark energy and dark matter. The coupling of the axionic dark matter to the non-uniformly moving aether was described in \cite{B16} based on the axionic extension of the Einstein-aether theory. The cosmological application of the Einstein-aether-axion theory shows explicitly, that the late-time Universe in this model expands with acceleration, and asymptotically, the Universe evolution is of the de Sitter type \cite{B16}. In other words, in the context of the Universe evolution the unit dynamic vector field, associated with the velocity four-vector of the aether motion, produces effects identified with the cosmic dark energy influence. In order to study the "dark properties" of the aether, one has to elaborate a set of physical tests clarifying the character of coupling  between the dynamic vector field and cosmic carriers of information. Keeping in mind this idea, we studied in \cite{BL14,AB15,2B16} a number of phenomenological models of interaction of the unit vector field with the electromagnetic field. When we deal with microscopic description of such interactions, we need to consider the dynamics of individual particles with spin or polarization. The coupling of relativistic spinning particles to the  axionic dark matter was described in \cite{BP15,BP16}; now we plan to study the influence of the dynamic aether on the relativistic particles with spin or polarization in the context of testing of the dark sector of the Universe.

The basic version of the Einstein-aether theory contains four Jacobson's coupling constants introduced phenomenologically, $C_1$, $C_2$, $C_3$, $C_4$ \cite{J1}. A physical sense of these parameters and their links with Newtonian gravitational constant, with post-Newtonian parameters, etc., were discussed in many works (see, e.g., the status report \cite{J7} for details and references).
For our discussion it is interesting to mention the following details. It is well known that the covariant derivative of the vector field $\nabla_iU_k$ can be decomposed into the sum of four irreducible parts; the first one contains the acceleration four-vector $DU_i$; the second item is the symmetric shear tensor $\sigma_{ik}$; the third element of the sum is the skew-symmetric vorticity tensor $\omega_{ik}$; the last addend contains the expansion scalar $\Theta$.
When one considers the application of the Einstein-aether theory to isotropic homogeneous cosmological models, only one effective coupling constant $C_{\theta}{=} C_1{+}3C_2{+}C_3$ enters the master equations \cite{J7}. In fact, three constants remain hidden due to the symmetry of such cosmological model, and the effective parameter $C_{\theta}$ appears in the Lagrangian in front of the squared  expansion scalar $\Theta^2$, which is the unique element of decomposition of the tensor $\nabla_iU_k$ in the model with such symmetry.
When one deals with static spherically symmetric models, the parameter $C_{\rm a}{=}C_1{+}C_4$ becomes the key parameter, which appears in front of the scalar $DU_k DU^k$. The Einstein-aether model with plane-wave symmetry involves into consideration the parameter $C_{\sigma}=C_1{+}C_{3}$, which appears in the Lagrangian in front of non-vanishing scalar $\sigma_{ik}\sigma^{ik}$. We are interested to consider the models, in which the skew-symmetric vorticity tensor $\omega_{ik}$ plays the main role, and the removal of degeneracy with respect to the effective coupling constant $C_{\omega} {=}C_1{-}C_3$ takes place. In other words, we intend now to consider the models, in which the rotational degree of freedom in the aether flow is activated.

One of the most known space-time models admitting the rotation in the velocity field, is the G\"odel space-time, which describes the universe rotation. The classical G\"odel solution \cite{Godel0} is obtained with the assumption that the universe contains a dust matter and is stabilized by the cosmological constant. One can say using modern terminology that this classical solution relates to the presence of a dark fluid (a dark matter of the dust type plus a dark energy of the $\Lambda$ type). There is a number of works (see, e.g. \cite{Godel1}--\cite{Godel16}), in which the universe rotation is associated with various modifications of the gravitational theory (non-minimal models, in particular $f(R)$, $f(T)$, $f(R,T)$ models, scalar-tensor theories, vector-tensor theories of gravity, etc.). The unit dynamic vector field, the basic element of the  Einstein-aether theory, was considered as a source of the universe rotation in the works \cite{Godel8,Godel15}.

Our goal is to study the behavior of  relativistic test particles possessing spin or polarization, which interact with dynamic unit vector field. Why this behavior seems to be interesting? Answering this question, we keep in mind the  analogy with the classical direct and reciprocal Magnus effects. When the rotating body moves in the fluid flow, it changes the direction of motion; when the fluid flow is characterized by non-vanishing circulation, the body changes the direction of motion. Now we deal with the aether flow, which possesses the rotational degree of freedom (vortex). Despite the relativistic particle is considered to be point-like, we can expect that similar Magnus-type interactions can appear in microscopic theory also, since the particle has a vectorial internal degree of freedom (spin or polarization). In this case the spin (polarization) four-vector attributed to the test particle, can be treated as a marker, and the spin precession or polarization rotation can signalize about the interaction of this particle with the non-uniformly moving aether. To describe the influence of the universe rotation on the spin (polarization) evolution we also use the extension of the G\"odel model based on the Einstein-aether theory. Our study includes some new details; for instance, we consider an exact solution of the G\"odel-type obtained for a matter substratum with non-vanishing anisotropic pressure (e.g., for the anisotropic dark fluid), as well as, we investigate  a new exact solution without matter, valid for the case, when the unit vector field with specific set of Jacobson's parameters is the source of the universe rotation.

The paper is organized as follows. In Section II, we remind the basic elements of the Einstein-aether theory and introduce the system of master equations for relativistic particle dynamics and spin (polarization) evolution accounting for interactions with the dynamic aether. In Section III, we consider the first application of the extended model: a G\"odel-type universe supported by aether with arbitrary Jacobson's constants and a dark fluid with anisotropic pressure; we discuss obtained exact solutions to the vector field equations, to the gravity field equations, to the equations of particle dynamics  and spin (polarization) evolution. In Section IV, we consider the second application of the extended model:  a G\"odel-type universe supported by a pure aether with one arbitrary Jacobson's constant. Section V contains discussion and conclusions.

\section{The formalism}
\label{principles}

\subsection{Action functional of the Einstein-aether theory}

The Einstein-aether theory is constructed using the action functional of the following form (see, e.g., \cite{J1}):
\begin{eqnarray}
S_{({\rm EA)}} &{=}&  \int  d^4 x \sqrt{{-}g}
\Bigl\{  \frac{1}{2\kappa} \left[R{+}2\Lambda {+} \lambda
\left(g_{mn}U^m U^n {-}1 \right) {+}\right. \Bigl.
\nonumber\\
&&
\Bigl. \left. {+} K^{abmn}   \nabla_a U_m
\nabla_b U_n \right] {+} L^{({\rm m})}  \!\Bigl\} \,. \label{S}
\end{eqnarray}
Here, the determinant of the metric $g {=} {\rm det}(g_{ik})$, the
Ricci scalar $R$, the cosmological constant $\Lambda$ are the standard elements
of the Einstein-Hilbert action. The term $L^{({\rm m})}$ is the
Lagrangian of a matter substratum.
Two new elements involving the vector field $U^i$ appear in
(\ref{S}). The first term, $\lambda \left(g_{mn}U^m U^n {-}1 \right) $,
ensures that the $U^i$ is normalized to one. The second term
contains all possible pair  convolutions of
covariant derivatives  of the vector field $U^i$. A necessary invariant structure is formed by Jacobson's constitutive tensor
$K^{abmn}$, which is constructed using the metric tensor
$g^{ij}$ and the velocity four-vector $U^k$ only:
\begin{equation}
K^{abmn} {=} C_1 g^{ab} g^{mn} {+} C_2 g^{am}g^{bn}
{+} C_3 g^{an}g^{bm} {+} C_4 U^{a} U^{b}g^{mn} .
\label{2}
\end{equation}
The parameters $C_1$, $C_2$, $C_3$ and $C_4$ are Jacobson's constants.

The tensor $\nabla_i U_k$ can be decomposed into a sum of
its irreducible parts: the acceleration four-vector $DU^{i}$,
the shear tensor $\sigma_{ik}$,
the vorticity tensor $\omega_{ik}$, and
the expansion scalar $\Theta$:
\begin{equation}
\nabla_i U_k = U_i DU_k + \sigma_{ik} + \omega_{ik} +
\frac{1}{3} \Delta_{ik} \Theta \,. \label{act3}
\end{equation}
The basic quantities are defined as follows:
\begin{eqnarray}
&\displaystyle
DU_k \equiv U^m \nabla_m U_k \,,
&\nonumber\\
&\displaystyle
\sigma_{ik}
\equiv \frac{1}{2}\Delta_i^m \Delta_k^n \left(\nabla_m U_n {+}
\nabla_n U_m \right) {-} \frac{1}{3}\Delta_{ik} \Theta  \,,
&\nonumber\\
&\displaystyle
\omega_{ik} \equiv \frac{1}{2}\Delta_i^m \Delta_k^n \left(\nabla_m
U_n {-} \nabla_n U_m \right) \,,
&\label{act4}\\
&\displaystyle
\Theta \equiv \nabla_m U^m
\,, \quad D \equiv U^i \nabla_i \,, \quad \Delta^i_k = \delta^i_k - U^iU_k  \,.
\nonumber
\end{eqnarray}
In these terms the scalar $K^{abmn}\nabla_a U_m \nabla_b U_n$ can be rewritten as
\begin{eqnarray}
&&
\!\!\!\!\!\!\!\!\!\!\!\!\!\!\!
K^{abmn}\nabla_a U_m \nabla_b U_n =
\nonumber\\&&
=(C_1 {+} C_4)DU_k DU^k +
(C_1 {+} C_3)\sigma_{ik} \sigma^{ik} +
\label{act5}\\&&
+  (C_1 {-} C_3)\omega_{ik}
\omega^{ik} + \frac13 \left(C_1 {+} 3C_2 {+}C_3 \right) \Theta^2
\,,
\nonumber
\end{eqnarray}
thus explaining the usage of the alternative set of coupling constants
\begin{eqnarray}
&\displaystyle
C_{\rm a} = C_1 {+} C_4 \,, \quad
C_{\sigma} = C_1 {+} C_3 \,,
&\nonumber\\
&\displaystyle
C_{\omega}=C_1 {-} C_3 \,, \quad
C_{\theta}=C_1 {+} 3C_2 {+}C_3 \,.
&\label{act59}
\end{eqnarray}
Clearly, the parameter $C_{\rm a}$ appears in the Lagrangian in front of the square of the acceleration four vector $a^i\equiv DU^i$, the parameter $C_{\omega}$ introduces the squared vorticity tensor, etc.

\subsection{Field equations}

\subsubsection{Master equations for the unit dynamic vector field}

The variation of the action (\ref{S}) with respect to
$\lambda$ yields the equation
\begin{equation}
g_{mn}U^m U^n = 1 \,,
\label{21}
\end{equation}
which is known to be the normalization condition of the time-like
vector field $U^k$. The variation of the functional (\ref{S}) with respect to vector field
$U^i$ gives the equations
\begin{equation}
\nabla_a {\cal J}^{aj} = I^j + \lambda \ U^j  \,,
\label{0A1}
\end{equation}
where
\begin{equation}
{\cal J}^{aj} = K^{abjn} \nabla_b U_n  \,, \quad
I^j =  C_4 (DU_m)(\nabla^j U^m) \,.
\label{U1}
\end{equation}
The Lagrange multiplier $\lambda$ can be found as
\begin{equation}
\lambda =  U_j \left[\nabla_a {\cal J}^{aj}- I^j \right]  \,,
\label{0A309}
\end{equation}
using the convolution of (\ref{0A1}) with the velocity four-vector.

\subsubsection{Master equations for the gravitational field}

The variation of the action (\ref{S}) with respect to the metric
$g^{ik}$ yields the gravitational field equations:
\begin{equation}
R_{ik} - \frac{1}{2} R \ g_{ik}  =  \Lambda g_{ik} + \lambda U_i
U_k  + T^{({\rm U})}_{ik} + \kappa T^{({\rm m})}_{ik}\,.
\label{0Ein1}
\end{equation}
The term $T^{({\rm U})}_{ik}$ is the stress-energy tensor of the vector field $U^i$:
\begin{eqnarray}
&&\!\!\!\!\!
T^{({\rm U})}_{ik} =
\frac12 g_{ik} {\cal J}^{am} \nabla_a U_m +
\nonumber\\&&
+\nabla^m \left[U_{(i}{\cal J}_{k)m} {-}
{\cal J}_{m(i}U_{k)} {-}
{\cal J}_{(ik)} U_m\right]+
\label{E11}\\&&
C_1\left[(\nabla_mU_i)(\nabla^m U_k) {-}(\nabla_i U_m)(\nabla_k U^m) \right] {+} C_4 D U_i D U_k \,,
\nonumber
\end{eqnarray}
where parentheses in the index line denote symmetrization, $p_{(i} q_{k)}{\equiv}\frac12 (p_iq_k{+}p_kq_i)$. The stress-energy tensor $T^{({\rm m})}_{ik}$ is
defined standardly as
\begin{equation}
T^{({\rm m})}_{ik} = -\frac{2}{\sqrt{-g}} \frac{\delta
[\sqrt{-g} L^{({\rm m})} ]}{\delta g^{ik}} \,.
\label{E13}
\end{equation}
Using the velocity four-vector of the aether $U^i$ this tensor can be algebraically decomposed  as
follows:
\begin{equation}
T^{({\rm m})}_{ik} = \rho U_i U_k + U_i I^{(q)}_k + U_k I^{(q)}_i +
P_{ik} \,.
\label{E12}
\end{equation}
Here $\rho$ is the energy density scalar with respect to a restframe of the aether, $I^{(q)}_k$ is the heat-flux
four-vector, the tensor $P_{ik}$ describes the anisotropic
pressure. These quantities are defined standardly as
\begin{eqnarray}
&\displaystyle
\rho = U^iT^{({\rm m})}_{ik} U^k \,, \quad I^{(q)}_p = U^i T^{({\rm
m})}_{ik} \Delta^{k}_p \,,
&\nonumber\\
&\displaystyle
P_{pq} = \Delta_p^{i} T^{({\rm
m})}_{ik} \Delta_q^{k} \,.
&\label{aaE12}
\end{eqnarray}
The four-vector $I^{(q)}_p$ and the symmetric tensor $P_{pq}$ are orthogonal to the four-vector $U^p$.

\subsection{Dynamics of particles with vectorial internal degree of freedom in the non-uniformly moving aether}
\label{sec - particle dynamics}

\subsubsection{Basic equations describing massive particles with spin}\label{sec - basic eqs}

The evolution of relativistic point-like particles with
an electric charge and a spin four-vector is guided by the following set of master equations (see, e.g.,  \cite{BK2004,BP15} for motivation and details):
\begin{equation}
\frac{D p^i}{D\tau} = {\cal F}^i \ , \quad \frac{D S^i}{D\tau} =
{\cal G}^i \,. \label{1}
\end{equation}
Here $p^{i}$ is the particle momentum four-vector, and $S^{i}$ is the spin four-vector.
The force-like terms, ${\cal F}^i $
and ${\cal G}^i $, describe the rates of change of the momentum and spin, respectively;
they possess  three general properties.

\noindent {\it (i)}. The mass of the particle, $m$, defined from
the normalization law $p^i p_i = m^2$ (we use the units with $\hbar{=}c{=}1$),  is assumed to be
conserved quantity, providing the four-vector ${\cal F}^i$ to be
orthogonal to the momentum:
\begin{equation}
 p_i {\cal F}^i  =  p_i \frac{D p^i}{D\tau} = \frac12\frac{D}{D\tau} (p_i p^i) =0
 \,.
\label{0s1}
\end{equation}

\noindent {\it (ii)}. Similarly, we assume that the scalar square
of the space-like spin four-vector is constant, i.e.,  $S^i S_i
=$const$= - {\cal S}^2$. Then using the second equation from
(\ref{1}) we obtain that
\begin{equation}
 S_i {\cal G}^i  =  S_i \frac{D S^i}{D\tau} = \frac12\frac{D}{D\tau} (S_i S^i)
 = - \frac12\frac{D}{D\tau} {\cal S}^2 =0
 \,,
\label{0s2}
\end{equation}
or in other words, that the force-like term ${\cal G}^i$ is
orthogonal to the spin four-vector.

\noindent {\it (iii)}. We assume that the spin
four-vector is orthogonal to the momentum four-vector, $S^ip_i=0$; then one obtains that
\begin{equation}
\frac{D}{D\tau} (p^i S_i) = 0 \ \ \Rightarrow \ {\cal F}^i  S_i +
{\cal G}_i p^i =0 \,. \label{111}
\end{equation}
There are two scalar consequences from the equations (\ref{1}): their convolutions with the velocity four-vector $U_i$ yield
\begin{eqnarray}
\frac{d}{d\tau} (p^i U_i) &=&   {\cal F}^i  U_i + \frac{1}{m} p^i p^j \nabla_{(i} U_{j)}
\,, \label{1d1}
\\
\frac{d}{d\tau} (S^i U_i) &=&   {\cal G}^i  U_i + \frac{1}{m} S^i p^j \nabla_{j} U_{i}
\,. \label{1d2}
\end{eqnarray}
When the covariant derivative $\nabla_{i} U_{j}$ is skew-symmetric, i.e., $\nabla_{(i} U_{j)}{=}0$, the equation (\ref{1d1}) shows that the particle energy ${\cal E} {=}p_iU^i$, calculated in the frame of reference associated with the unit vector field $U^i$, can be changed by the force ${\cal F}^i$ only.

Three equations (\ref{0s1}), (\ref{0s2}) and (\ref{111}) are satisfied identically, when
\begin{equation}
{\cal F}^i  = \Omega^{ik} p_k  \,, \quad {\cal G}^i = \Omega^{ik}
S_k \,, \label{BMT1}
\end{equation}
where $\Omega^{ik}$ is arbitrary skew-symmetric tensor,
$\Omega^{ik}{=} {-} \Omega^{ki}$.
In the minimal Einstein-aether theory the quantities ${\cal F}^i $ and ${\cal G}^i$
depend on the particle momentum $p^k$, spin $S^l$, velocity four-vector $U^i$, acceleration four-vector $DU^i$,
shear tensor $\sigma_{ik}$, vorticity tensor $\omega_{ik}$ and expansion scalar $\Theta$.
In the nonminimally extended theory the
quantities ${\cal F}^i $ and ${\cal G}^i $ can include also the Riemann
tensor $R^{i}_{\ klm}$, Ricci tensor $R_{ik}$, Ricci scalar
$R$, and the covariant derivative of the Riemann tensor $\nabla_s R_{ikmn}$ (see, e.g., \cite{BK2004}).

\subsubsection{Basic equations for massless particles with polarization}

When we are interested to consider test massless particles  ($m{=}0$), the internal vectorial degree of freedom is associated with the polarization four-vector $\xi^i$, which is assumed to be space-like and normalized by unity ($\xi^i \xi_i =  {-}1$). In this case we have to use the four-vector $k^i$ instead of $p^i$, and have to replace $\frac{p^i}{m}$ by $k^i$ in all the equations of particle dynamics. The normalization condition reads now as $k^i k_i =0$. The parameter $\tau$ along the particle world-line is now defined so that $k^i {=} \frac{dx^i}{d\tau}$. For the massless particles the basic equations transform into
\begin{equation}
\frac{D k^i}{D\tau} = {\cal F}^i \ , \quad \frac{D \xi^i}{D\tau} =
{\cal G}^i \,, \label{j1}
\end{equation}
providing the auxiliary relations
\begin{eqnarray}
&\displaystyle
 k_i {\cal F}^i  = 0 \,, \quad  \xi_i {\cal G}^i  = 0 \,, \quad {\cal F}^i  \xi_i +
{\cal G}_i k^i =0  \,,
\label{j0s1}
&\\
&\displaystyle
\frac{d}{d\tau} (k^i U_i) =   {\cal F}^i  U_i +  k^i k^j \nabla_{(i} U_{j)}
\,, \label{01d1}
&\\
&\displaystyle
\frac{d}{d\tau} (\xi^i U_i) =   {\cal G}^i  U_i + \xi^i k^j \nabla_{j} U_{i}
\,. \label{01d2}
&
\end{eqnarray}
The reconstruction of the force-like terms  ${\cal F}^i$ and ${\cal G}^i$ gives two principally new details in comparison with the case of massive particle.
The first novelty is that one can add the terms proportional to the null four-vector $k_i$ to both quantities ${\cal F}^i$ and ${\cal G}^i$:
\begin{equation}
{\cal F}^i  = \Omega^{ij} k_j + \mu k^i  \,, \quad {\cal G}^i = \Omega^{ij}
\xi_j + \tilde{\mu} k^i \,, \label{jBMT1}
\end{equation}
providing the relations (\ref{j0s1}) to be satisfied, since $k_ik^i=0$ and $\xi_i k^i=0$. The scalar $\mu$ can be arbitrary function of the scalar ${\cal E}{=}k_iU^i$ only. As for the
scalar $\tilde{\mu}$, it is considered to be a linear function of the scalar $\xi_i U^i$ and arbitrary function of the particle energy  ${\cal E}$.

The second distinction from the case of massless particle is that now there is some kind of gauge degree of freedom in the definition of the polarization four-vector. Indeed, if we put
\begin{equation}
\tilde{\xi}_i = \xi_i + {\cal R} k_i \,,
\label{jmm2}
\end{equation}
we guarantee that the same normalization-orthogonality conditions
\begin{equation}
\tilde{\xi}_i \tilde{\xi}^i = -1 \,, \quad \tilde{\xi}_i k^i = 0   \label{kq1}
\end{equation}
are satisfied for the transformed polarization four-vector. After the transformation (\ref{jmm2}), the first equation in (\ref{j1}) remains unchanged, the second equation keeps the form
\begin{equation}
\frac{D \tilde{\xi}_i}{D\tau} = \Omega_{il} \tilde{\xi}^l + k_i \left[\tilde{\mu} + {\cal R} \mu + \frac{d{\cal R}}{d\tau} \right] \,. \label{j1mm}
\end{equation}
There are two interesting consequences from the last equation.

\noindent
{\it (j)} When  $\frac{d{\cal R}}{d\tau} {=} {-} {\cal R} \mu $ the transformation (\ref{jmm2}) retains the the equation for polarization unchanged.

\noindent
{\it (jj)} When  $\tilde{\mu} {+} {\cal R} \mu {+} \frac{d{\cal R}}{d\tau} {=}0$  the transformation (\ref{jmm2}) excludes the term linear in $k_i$ from the equation of the polarization evolution.
In particular, if $\mu{=}0$ (it is the physically motivated choice, which guarantees that the massless particle energy remains constant in the absence of external impacts) the scalar ${\cal R}$ can be found from the condition $\frac{d{\cal R}}{d\tau} {=} {-} \tilde{\mu}$ to redefine the polarization four-vector. We will use such a gauge transformation of the polarization four-vector below in Subsection IIIE3.

\subsubsection{Reconstruction of the force-type sources}\label{sec -
reconstruction}

Keeping in mind the general relationships (\ref{BMT1}) we
reconstruct the tensor $\Omega^{ik}$ using the following ansatz:

\noindent
A) the tensor   $\Omega^{ik}$ contains neither $S^i$, nor $\xi^i$;

\noindent
B) the tensor $\Omega^{ik}$ is linear in the covariant
derivative of the unit vector field.

The requirements A)  is based on the idea that in the leading order of semi-classical approach the Planck constant $\hbar$ does not enter the master equations; since the spin four-vector appears in this approach in the form of product $\hbar S^i$, it seems to be reasonable to eliminate such a quantity from $\Omega^{ik}$.
The requirement B) is typical for forces of different kind. This approach is used in the field theory, in transport phenomena, etc., and imply that the forces are caused by inhomogeneous potentials, flows, etc.

In order to simplify the irreducible decomposition of the tensor
$\Omega^{ik}$ we use the representation
\begin{equation}
p^i =U^i {\cal E} + q^i \,, \quad k^i = U^i {\cal E} + q^i \,,
\label{vv1}
\end{equation}
where the particle energy is determined with respect to the aether velocity ${\cal E} \equiv U^k p_k $ or ${\cal E} \equiv U^j k_j $, and the transversal component of the momentum four-vector is defined as
\begin{equation}
q^i \equiv \Delta^i_k p^k = p^i
- {\cal E} U^i \,.
\label{vv2}
\end{equation}
Also, we assume that the coefficients in the decomposition can be
appropriate functions of the scalar ${\cal E}$. The decomposition
contains terms of zero order with respect to $q^i$, linear,
quadratic etc. For instance, zero-order and linear terms in $q^i$ can be written
as follows:
\begin{eqnarray}
&&\!\!\!\!\!\!\!
\Omega_{ik} = \frac12 \delta^{mn}_{ik}\Bigl[\alpha_1 U_m DU_n + \alpha_2
\omega_{mn}  + q_m U_n
\left(\alpha_3 + \alpha_4 \Theta \right) + \Bigl.
\nonumber\\&&\qquad
\Bigl.   + q^j U_m \left(\alpha_5
\sigma_{nj}+ \alpha_6 \omega_{nj}\right) + \alpha_7 q_m D U_n
\Bigl] \,. \label{OmegaForm}
\end{eqnarray}
In this approximation the model contains seven coupling parameters, when $m \neq 0$, and nine coupling constants, when $m=0$ (two additional parameters can appear when $\mu \neq 0$ and $\tilde{\mu} \neq 0$).

\section{First application: G\"odel-type universe supported by aether with arbitrary Jacobson's constants}

\subsection{The space-time metric}

The metric describing the rotating universe appeared in the original work \cite{Godel0}.
Using the formal redefinitions of the coordinates ($ax_0 \to t$, $ax_1 \to x$, etc.), and of the parameter $a \to \frac{1}{\nu}$, we rewrite the G\"odel metric as follows:
\begin{equation}
ds^2 = dt^2 - dx^2 + \frac12 e^{2\nu x}dy^2 + 2e^{\nu x}dtdy -
dz^2 \,.
\label{G1}
\end{equation}
The constant $\nu$ is connected with the Ricci scalar calculated for this metric as $R {=} \nu^2$.
When $\nu=0$, the space-time is flat, and we can return to the Minkowski metric using the coordinate transformation $\tilde{t}=t+y$, $\tilde{y}=\frac{y}{\sqrt2}$.
In the original work \cite{Godel0} the author assumed that the velocity four-vector, associated with the source of the gravitational field, has only one non-vanishing contravariant
component $u^i {=}(\frac{1}{a},0,0,0)$. Since in that case the covariant components of the Ricci tensor, $R_{ik}$, happened to be proportional to the product $u_iu_k$, this assumption
led the author of \cite{Godel0} to the conclusion that the source of the corresponding gravitational field is the dust with constant energy-density, and the cosmological constant, the sign of which is opposite of that occurring in the Einstein static solution.

Below we consider two models of the Einstein-aether type. In the first one we use the G\"odel's ansatz about the velocity four-vector; the
corresponding exact solutions contain arbitrary Jacobson's constants. In the second model we considered the velocity four-vector with two non-vanishing components; the corresponding exact solutions   are found for Jacobson's constants coupled by three relationships.

\subsection{Ansatz about the structure of the unit dynamic vector field}

In the model under consideration, for the unit dynamic  vector field we use the ansatz, which was proposed by G\"odel in \cite{Godel0}:
\begin{equation}
U^i = \delta^i_0  \ \ \ \Rightarrow  U_k = \delta_k^0 + \delta_k^2
e^{\nu x}\,.
\label{G2}
\end{equation}
The new detail is that in \cite{Godel0} the four-vector $u^i$ was the eigen-vector of the stress-energy tensor of the dust, while in our approach we deal with the
unit dynamic  vector field $U^i$ as one of the sources of the gravity field. Another source in our model is a Dark Fluid with an anisotropic pressure.

The covariant derivative $\nabla_iU_k$ can be now
written as
\begin{equation}
\nabla_iU_k = \frac12 \nu  e^{\nu x}\left(\delta_i^1 \delta_k^2
-\delta_i^2\delta_k^1 \right) \,,
\label{G3}
\end{equation}
providing that the acceleration four-vector, the shear tensor and the expansion scalar vanish
\begin{equation}
DU_k =0 \,, \quad \sigma_{ik}=0 \,, \quad \Theta =0 \,.
\label{G4}
\end{equation}
Only the vorticity tensor is non-vanishing
\begin{equation}
\omega_{ik} = \nabla_iU_k = \frac12 \nu e^{\nu x}\left(\delta_i^1
\delta_k^2 -\delta_i^2\delta_k^1 \right) \,.
\label{G5}
\end{equation}
The square of the vorticity tensor is constant
\begin{equation}
\omega^{pq} \omega_{pq} = \nu^2 \,.
\label{G6}
\end{equation}
The corresponding angular velocity four-vector is
\begin{equation}
 \omega^{j} \equiv \omega^{*jn}U_n = \frac12 \omega_{pq} \epsilon^{jnpq} U_n = - \frac{\nu}{\sqrt2}
\delta^j_3  \,,
\label{G7}
\end{equation}
where $\epsilon^{jnpq}{=}\frac{E^{jnpq}}{\sqrt{-g}}$ is the Levi-Civita tensor, $E^{jnpq}$ is the absolutely anti-symmetric Levi-Civita symbol with $E^{0123}=+1$. Clearly, the parameter $\nu$ describes the angular velocity of the universe uniform rotation, since the observer moving with the velocity $U^i$ is in the rest with respect to the universe.

\subsection{Exact solutions to the field equations}

\subsubsection{Solutions to the reduced equations  for the unit dynamic vector field}

Since only the vorticity tensor is non-vanishing in this model, we obtain using (\ref{G3}) that
\begin{equation}
I^j =0 \,, \quad {\cal J}^{aj}=  - {\cal J}^{ja} =
\left(C_1-C_3\right) \omega^{aj} \,.
\label{G10}
\end{equation}
Thus, the field equations (\ref{0A1}) are  transformed into
\begin{equation}
\left(C_1-C_3\right) e^{-\nu x} \frac{d}{dx} \left[e^{\nu x}
\omega^{1j} \right] = \lambda \ U^j  \,, \label{G11}
\end{equation}
and yield only one non-trivial equation
\begin{equation}
-\nu^2 \left(C_1-C_3\right)  = \lambda \,. \label{G118}
\end{equation}
In other words, the equations for the velocity four-vector happen
to be satisfied for arbitrary Jacobson's parameters $C_1$, $C_2$,
$C_3$, $C_4$, and the Lagrange multiplier $\lambda$ is constant and is proportional to $C_{\omega}{=}C_1 {-} C_3$.

\subsubsection{Solutions to the reduced equations for the gravity field}

The stress-energy tensor of the unit vector field takes now the
form
\begin{equation}
T^{({\rm U})}_{ik} {=}
C_{\omega} \left[\frac12 g_{ik}
\omega^{am} \omega_{am} {+} \nabla^m \left[U_{i}\omega_{km} {+}
U_{k}\omega_{im}\right] \right] \,.
\label{G14}
\end{equation}
The gravity field equations
\begin{eqnarray}
&&\!\!\!\!\!\!\!
R_{ik} - \frac{1}{2} R \ g_{ik}  -  \Lambda g_{ik} - \kappa
T_{ik}^{({\rm m})} =
\nonumber\\&&
= \nu^2 C_{\omega}
\left[\frac{3}{2} g_{ik} {+} 2\delta^1_i \delta^1_k {+}  e^{2\nu x}
\delta^2_i\delta^2_k {+} \delta^3_i \delta^3_k \right]
\label{G17}
\end{eqnarray}
give five nontrivial equations for the metric (\ref{G1})
\begin{eqnarray}
\frac12 \nu^2 - \Lambda - \kappa T_{00}^{({\rm m})}
&=&
\frac32\nu^2 C_{\omega} \,,
\nonumber\\
\frac12 \nu^2  e^{\nu x} - \Lambda e^{\nu x} - \kappa T_{02}^{({\rm m})}
&=&
\frac32 \nu^2 C_{\omega} e^{\nu x} \,,
\nonumber\\
\frac12 \nu^2   + \Lambda - \kappa T_{11}^{({\rm m})}
&=&
\frac12 \nu^2 C_{\omega} \,,
\label{G18}\\
\frac34 \nu^2 e^{2\nu x} - \Lambda \frac12 e^{2\nu x} - \kappa T_{22}^{({\rm m})}
&=&
\frac74 \nu^2 C_{\omega} e^{2\nu x} \,,
\nonumber\\
\frac12 \nu^2    + \Lambda - \kappa T_{33}^{({\rm m})}
&=&
-\frac12 \nu^2 C_{\omega} \,.
\nonumber
\end{eqnarray}
Compatibility conditions based on the Bianchi identities are reduced
to one equation
\begin{equation}
\left[\frac{d}{dx} + \nu \right] T_{11}^{({\rm m})} + 2\nu
\left[e^{-\nu x} T_{02}^{({\rm m})} -  e^{-2\nu x}
T_{22}^{({\rm m})} \right] =0 \,.
\label{G19}
\end{equation}
We assume that the matter substratum is presented by some quasi-dust matter, which has the pressure only in the longitudinal direction, i.e.,
in the direction of the rotation axis:
\begin{equation}
T_{ik}^{({\rm m})} = \rho U_i U_k + \delta_i^3 \delta_k^3 \Pi
\,.
\label{G20}
\end{equation}
Clearly, the compatibility conditions (\ref{G19}) are satisfied identically for the tensor (\ref{G20}). The gravity field equations are satisfied, when
\begin{eqnarray}
&\displaystyle
\kappa \Pi = \nu^2 C_{\omega} \,,
\label{G23}
&\\
&\displaystyle
\Lambda = \frac12 \nu^2 \left[ C_{\omega} -1 \right] \,,
\label{G26}
&\\
&\displaystyle
\kappa \rho = \nu^2 \left[1- 2C_{\omega} \right] \,,
\label{G27}
\end{eqnarray}
and the coupling constants $C_1$, $C_2$, $C_3$, $C_4$ are arbitrary. Thus, we obtained the
exact solution for the total self-consistent system of equations, for
which the unit vector field can be considered as the source of the universe rotation, and some material substratum with constant longitudinal pressure supports the system as a whole to be stationary.

\subsubsection{Remark concerning the specific case $C_1{=}C_3$}

When $C_1{=}C_3$, i.e., $C_{\omega}{=}0$, the vorticity of the aether flow exists, but it remains hidden in the master equations for both the unit vector field and the gravitational field. In other words, when $C_1{=}C_3$, the aether is insensible to perturbations of the vorticity type. The coupling parameters $C_{\rm a}$, $C_{\sigma}$, $C_{\theta}$ are non-vanishing, however, they are not activated, since the acceleration, shear and expansion of the aether flow are absent. Finally, when $C_1=C_3$, the obtained solution formally coincides with the known G\"odel's solution with a dust, since for this case
\begin{equation}
\Lambda = - \frac12 \nu^2 \,, \quad \kappa \rho = \nu^2 \,, \quad
\kappa \Pi =0 \,.
\label{G28}
\end{equation}
However, we would like to stress that it is an exact solution of the Einstein-aether model for the  arbitrary coupling constants $C_{\rm a}$, $C_{\sigma}$, $C_{\theta}$.

\subsection{Solutions to the equations of the spinning particle dynamics}

\subsubsection{Reduced master equations}\label{spd in GASu}

Since only the vorticity tensor $\omega_{ik}$ appears in the decomposition of the covariant derivative of the velocity four-vector $\nabla_iU_k$, many coefficients in (\ref{OmegaForm}) happen to be hidden (e.g., $\alpha_1$, $\alpha_4$, $\alpha_5$, $\alpha_7$). In order to study an appropriate example, which demonstrates the typical behavior of a spinning particle in this background we restrict ourselves by the following ansatz concerning  the tensor $\Omega_{ik}$:
\begin{equation}
\Omega_{ik} =  \frac{\alpha}{m} \ \omega_{ik} \,.
\label{OF2}
\end{equation}
Thus, in this section we deal with the following set of evolutionary equations describing the behavior of the spinning particle:
\begin{eqnarray}
\frac{Dp_i}{D\tau} &=& \frac{\alpha}{m}   \omega_{ik} p^k \,,
\label{OF3}\\
\frac{DS_i}{D\tau} &=& \frac{\alpha}{m}  \omega_{ik} S^k  \,.
\label{OF3s}
\end{eqnarray}
Here we introduced a new constant $\alpha = m \alpha_2$, since in this terms the equations (\ref{OF3}), (\ref{OF3s}) are similar to the Bargmann--Michel--Telegdi equations without abnormal magnetic moment (see, e.g., \cite{BMT}). In the equations (\ref{OF3}), (\ref{OF3s}) the term $\omega_{ik}$ plays the same role, which the Maxwell tensor $F_{ik}$ plays in the Lorentz force, so the constant $\alpha$ has a sence of an effective charge. The force ${\cal F}_i$ itself is the analog of the Lorentz force, but the origin of this force is different: it is produced by the rotation of the aether flow.

\subsubsection{Particle dynamics}

We consider, first, the equations of the particle dynamics using the representation (\ref{G5}) of the vorticity tensor. These equations can be written as follows:
\begin{equation}
\frac{dp_i}{d\tau} - \frac{1}{2m} p^j p^k \partial_i g_{jk} =
\frac{1}{2m} \nu \alpha e^{\nu x} \left(\delta_i^1 p^2 -\delta_i^2
p^1 \right) \,.
\label{OF4}
\end{equation}
Since the metric does not depend on time, we obtain immediately that
\begin{equation}
\frac{dp_0}{d\tau} = 0  \ \Rightarrow  \ p_0 (\tau) = K_0 \,,
\label{OF5}
\end{equation}
i.e., the particle energy ${\cal E}=U^ip_i=p_0=K_0$ is constant along the world-line.
This result can also be immediately obtained from the equation (\ref{1d1}), since the symmetric part of the covariant derivative $\nabla_{(i} U_{j)}$ vanishes, and
${\cal F}_i  U^i {=} {\cal F}_0 {=0}$.
Similarly, we obtain that
\begin{equation}
\frac{dp_3}{d\tau} = 0 \ \Rightarrow \ p_3 = K_3 \,,
\label{OF6}
\end{equation}
i.e., the longitudinal component (with respect to the universe rotation axis) of the momentum four-vector is constant.
In order to find $p_1(\tau)$ and $p_2(\tau)$ we use the following procedure. First, we
assume that the component $p^1$ of the particle momentum is non-vanishing, and thus, we can link the parameter
$\tau$ along the particle world-line with the variable $x$ by the relationship
\begin{equation}
\frac{d}{d\tau} = \frac{p^1}{m} \frac{d}{dx} \,.
\label{OF7}
\end{equation}
Second, using the variable $x$ we solve the equation for $p_2$
\begin{equation}
\frac{dp_2}{dx} = - \frac12 \nu \alpha e^{\nu x}  \
\Rightarrow p_2(x) = K_2 - \frac12 \alpha e^{\nu x} \,.
\label{OF8}
\end{equation}
Finally, the equation for $p_1$
\begin{equation}
\frac{dp_1}{d\tau} - \frac{1}{2m}\left[ 2p^0 p^2  g^{\prime}_{02}
+ {p^2}^2  g^{\prime}_{22}\right] = \frac{1}{2m} \nu \alpha e^{\nu x}
p^2 \,,
\label{OF9}
\end{equation}
can be resolved using the normalization condition $m^2 {=} g^{ik}p_i
p_k$ written as follows:
\begin{eqnarray}
&&\!\!\!\!\!\!
p^2_1  = 2 K_0 \left(2K_2 e^{{-}\nu x} {-} \alpha \right) -
\nonumber\\&&
 - m^2 - K_0^2 - K^2_3 - \frac12 \left(2 K_2 e^{{-}\nu x} {-} \alpha \right)^2 \,.
\label{OF10}
\end{eqnarray}
The relation between $\tau$ and $x$ can be found in formal
quadratures
\begin{equation}
\tau-\tau_{*} = m \int \frac{dx}{p_{1}(x)} \,,
\label{OF11}
\end{equation}
where $\tau_{*}$ is an integration constant.
The integration in this relationship requires a more detailed
analysis and we will return to this problem below.

\subsubsection{Physical components of the particle momentum}

As usual, to interpret physically the solutions for the particle momentum and spin we consider the projections of the corresponding four-vectors onto the tetrad four-vectors $X^i_{(a)}$,
which satisfy the relations
\begin{equation}
g_{ik} X^i_{(a)} X^k_{(b)} = \eta_{(a)(b)} \,, \quad \eta^{(a)(b)}
X^i_{(a)} X^k_{(b)} = g^{ik}\,,
\label{G8}
\end{equation}
with the symbol $\eta_{(a)(b)}$ denoting the Minkowski tensor.
For the G\"odel space-time the tetrad components have the form
\begin{eqnarray}
&\displaystyle
X^i_{(0)} = U^i =  \delta^i_{0} \,, \quad X^i_{(1)} = \delta^i_{1}
\,,
&\nonumber\\
&\displaystyle
X^i_{(2)} = \sqrt2\left(e^{{-}\nu x} \delta^i_{2} {-}
\delta^i_{0} \right) \,, \quad X^i_{(3)} = \delta^i_{3} \,.
\label{G9}&
\end{eqnarray}
The physical components of the momentum are
\begin{equation}
\Pi_{(a)} = X^i_{(a)} p_i \,,
\label{OF12}
\end{equation}
and using the formulas (\ref{G9}) we obtain the following three quantities:
\begin{eqnarray}
&\displaystyle
\Pi_{(0)}(x)  = X^i_{(0)} p_i = K_0 = \Pi_{(0)}(0) \equiv {\cal E} \,,
&\nonumber\\
&\displaystyle
\Pi_{(3)}(x) = X^i_{(3)} p_i = K_3 = \Pi_{(3)}(0) \equiv
\Pi_{||}\,,
\label{OFqq}&\\
&\displaystyle
\Pi_{(2)}(x) = X^i_{(2)} p_i = \sqrt2 \left[e^{-\nu x} K_2 -
\left(K_0+\frac12 \alpha \right) \right] \,.
&\nonumber
\end{eqnarray}
Using the value $\Pi_{(2)}(0)$ instead of parameter $K_2$ and ${\cal E}$ instead of $K_0$ we obtain the convenient formula for $\Pi_{(2)}(x)$
\begin{equation}
\Pi_{(2)}(x) = e^{{-}\nu x} \Pi_{(2)}(0) {+} \sqrt2 \left(e^{{-}\nu x}{-}1
\right) \left({\cal E}{+}\frac12  \alpha \right)  \,.
\label{OF14}
\end{equation}
In order to present $\Pi_{(1)}(x) {=} X^i_{(1)} p_i $ in an appropriate form, we attract the attention to the following fact:
\begin{equation}
\frac{d}{dx}\left[\Pi^2_{(1)}(x)+ \Pi^2_{(2)}(x)  \right]= 0 \,.
\label{p1}
\end{equation}
This means that the sum of squares of the transversal physical components of the particle momentum is constant:
\begin{eqnarray}
\Pi^2_{\bot} \equiv  \Pi^2_{(1)}(x) + \Pi^2_{(2)}(x)=
\Pi^2_{(1)}(0) + \Pi^2_{(2)}(0) =
&&\nonumber\\=
K_0^2 -m^2-K_3^2 = \text{const} \,.
\label{OF15}
\end{eqnarray}
This fact allows us to introduce the auxiliary convenient formulas
\begin{equation}
\Pi_{(1)}(x)=\Pi_{\bot} \cos{\Psi(x)} \,, \quad
\Pi_{(2)}(x)=\Pi_{\bot} \sin{\Psi(x)} \,.
\label{OF16}
\end{equation}
These formulas give the links between the parameter $\tau$ along the particle world-line  and the variable $x$.

\subsubsection{Links between $\tau$  and $x$}

The relation $\frac{dx}{d\tau} = \frac{p^1(x)}{m}$ yields the integral
\begin{equation}
\frac{(\tau-\tau_{*})}{m} = -\int \frac{dx}{\Pi_{(1)}(x)}\,,
\label{sfOF18}
\end{equation}
which can be transformed as follows:
\begin{equation}
\pm \frac{\nu (\tau-\tau_{*}) \Pi_{\bot}}{m} = \int \frac{d\Psi}{a
+ \sin{\Psi}}\,.
\label{OF18}
\end{equation}
Here the multiplier $\pm 1$ relates to positive and negative values of $p^1$, respectively,  and the guiding parameter $a$ is defined as
\begin{equation}
a \equiv \sqrt2 \ \frac{({\cal E} + \frac12 \alpha) }{\Pi_{\bot}} \,.
\label{OF19}
\end{equation}
The final result of integration in (\ref{OF18}) can be restructured  depending  on the value of the parameter $a$.
Let us mention, that when $\alpha \geq 0$, we obtain $a>1$. When $\alpha < 0$,
there are three intrinsic cases: $|a|>1$, $|a|= 1$, $|a|<1$. Let us consider these cases in more details.

\vspace{3mm}
\noindent {\it (i)} $|a|>1$

\noindent
The integration in (\ref{OF18}) gives
\begin{equation}
\tan{\frac{\Psi}{2}} = \pm \frac{\sqrt{a^2{-}1}}{a}
\tan{\left[\frac{\nu (\tau{-}\tau_{*}) \Pi_{\bot}}{2m}
\sqrt{a^2{-}1}\right]} {-}\frac{1}{a} \,.
\label{OF21}
\end{equation}

\vspace{3mm}
\noindent {\it (ii)} $a = \pm 1$

\noindent
We obtain from (\ref{OF18}) that
\begin{equation}
\tan{\frac{\Psi}{2}} = \mp \left[\frac{2m}{\nu (\tau-\tau_{*})
\Pi_{\bot}}\right] \mp 1 \,.
\label{OF22}
\end{equation}

\vspace{3mm} \noindent {\it (iii)} $|a| <1$

\noindent
The result of integration is
\begin{eqnarray}
&\displaystyle
\tan{\frac{\Psi}{2}} = - \frac{1}{a}\left[1+\sqrt{1-a^2}
\tanh^{-1}{\Gamma}\right] \,,
&\nonumber\\
&\displaystyle
 \Gamma \equiv \pm \frac{\nu (\tau-\tau^{*}) \Pi_{\bot}
\sqrt{1-a^2}}{2m} \,.
\label{OF23}&
\end{eqnarray}
When the parameter $\tau$ is expressed in terms of the quantity $\tan{\frac{\Psi}{2}}$, we can link, finally, $\tau$ and $x$, e.g.,
using the formula
\begin{eqnarray}
e^{-\nu x} \Pi_{(2)}(0) + \sqrt2 \left(e^{-\nu x}-1 \right) \left({\cal
E}+\frac12 \alpha \right) =
&&\nonumber\\
= 2 \Pi_{\bot} \left[
\frac{\tan{\frac{\Psi}{2}}}{1+\tan^2{\frac{\Psi}{2}}} \right]
\label{OF20}&&
\end{eqnarray}
obtained for $\Pi_{(2)}(x)$ (see (\ref{OF14}) and (\ref{OF16})).

\subsubsection{Evolutionary equations for the spin four-vector}

For the metric (\ref{G1}) the equations (\ref{OF3s}) take the form
\begin{eqnarray}
\frac{dS_i}{d\tau} - \frac{1}{2m} S^j p^k \left[\delta_k^1
g^{\prime}_{ij}+ \delta_i^1 g^{\prime}_{kj} - \delta_j^1
g^{\prime}_{ik}\right] =
&&\nonumber\\
=\frac{1}{2m} \alpha \nu e^{\nu x}
\left(\delta_i^1 S^2 -\delta_i^2 S^1 \right)\,.
\label{s1}&&
\end{eqnarray}
In order to simplify these equations we use the tetrad components of the spin four-vector $S_{(a)} \equiv X^i_{(a)} S_i$:
\begin{eqnarray}
&\displaystyle
S_{(0)} = S_0 \,, \quad S_{(1)} = S_1 \,, \quad S_{(3)} = S_3\,,
&\nonumber\\
&\displaystyle
S_{(2)} = \sqrt2 \left[e^{-\nu x} S_2 - S_0 \right] \,.
\label{s2}&
\end{eqnarray}
In these terms the orthogonality-normalization conditions
\begin{eqnarray}
&&g^{ik}p_i S_k =0 \qquad  \Rightarrow \
\nonumber\\
&&-{\cal E} S_{(0)} + \Pi_{(1)}
S_{(1)} + \Pi_{(2)} S_{(2)} +\Pi_{(3)} S_{(3)} =0 \,,
\label{s3}\\
&&g^{ik}S_i S_k = - {\cal S}^2 \qquad  \Rightarrow \
\nonumber\\
&&
{\cal S}^2 =
S^2_{(1)} + S^2_{(2)} + S^2_{(3)} - S^2_{(0)} \,,
\label{s4}
\end{eqnarray}
do not depend explicitly on the metric coefficients.
Using the same relationship between $\tau$ and $x$ as for the dynamic equations (see (\ref{OF18})--(\ref{OF20})), we can rewrite the equations for the spin evolution as follows:
\begin{eqnarray}
\frac{dS_{(0)}}{dx} & =& \frac{\nu }{\sqrt2 \Pi_{(1)}}  \left[ S_{(1)} \Pi_{(2)} - S_{(2)} \Pi_{(1)} \right] \,,
\label{s5}\\
\frac{dS_{(1)}}{dx} &=& \frac{\nu}{\sqrt2 \Pi_{(1)}}\biggl\{ S_{(0)} \Pi_{(2)} + \biggl.
\nonumber\\&&
\biggl.\qquad\qquad +S_{(2)}\left[\alpha {+} \sqrt2 \Pi_{(2)} {+} \Pi_{(0)} \right] \biggl\} \,,
\label{s7}\\
\frac{dS_{(2)}}{dx} &=& {-}\frac{\nu}{\sqrt2 \Pi_{(1)}}\biggl\{ S_{(0)} \Pi_{(1)} + \biggl.
\nonumber\\&&
\biggl. \qquad\qquad +S_{(1)}\left[\alpha {+} \sqrt2 \Pi_{(2)} {+} \Pi_{(0)} \right] \biggl\}\,,
\label{s8}\\
\frac{dS_{(3)}}{dx} &=& 0  \,.
\label{s6}
\end{eqnarray}
The last equation gives $S_{(3)}=E_{(3)}=$const, so that the longitudinal component of spin four-vector remains constant and is decoupled from the key set of evolutionary equations (\ref{s5})--(\ref{s8}) for $S_{(0)}$, $S_{(1)}$, $S_{(2)}$.

The solutions to the set of the equations (\ref{s5})--(\ref{s8}) with integration constants associated with relationships (\ref{s3}) and (\ref{s4}) can be presented in the following form:
\begin{eqnarray}
&&\!\!\!\!\!\!\!\!\!\!\!\!
S_{(0)}(x) = S_{(3)} \ \frac{\Pi_{(0)}\Pi_{(3)}}{(\Pi^2_{(0)}{-}\Pi^2_{\bot})} + {\cal A} \Pi^2_{\bot}  \cos{\Phi(x)}\,,
\label{s15}\\
&&\!\!\!\!\!\!\!\!\!\!\!\!
S_{(1)}(x) = S_{(3)} \ \frac{\Pi_{(1)}\Pi_{(3)}}{(\Pi^2_{(0)}{-}\Pi^2_{\bot})} +
\nonumber\\&&\!\!\!\!\!\!\!\!\!
{+}{\cal A}\left[\Pi_{(0)}\Pi_{(1)}\cos{\Phi(x)} {-} \Pi_{(2)} \sqrt{\Pi^2_{(0)}{-}\Pi^2_{\bot}}\sin{\Phi(x)} \right],
\label{s16}\\
&&\!\!\!\!\!\!\!\!\!\!\!\!
S_{(2)}(x) = S_{(3)} \ \frac{\Pi_{(2)}\Pi_{(3)}}{(\Pi^2_{(0)}{-}\Pi^2_{\bot})} +
\nonumber\\&&\!\!\!\!\!\!\!\!\!
{+} {\cal A}
\left[\Pi_{(0)}\Pi_{(2)}\cos{\Phi(x)} {+} \Pi_{(1)} \sqrt{\Pi^2_{(0)}{-}\Pi^2_{\bot}}\sin{\Phi(x)} \right].
\label{s17}
\end{eqnarray}
In these formulas the quantity  ${\cal A}$ given by
\begin{equation}
{\cal A} \equiv  \frac{\sqrt{(\Pi^2_{(0)}{-}\Pi^2_{\bot}) \ {\cal S}^2-m^2 S^2_{(3)}}}{\Pi_{\bot}(\Pi^2_{(0)}{-}\Pi^2_{\bot})}
\,,
\label{defA}
\end{equation}
contains the integrals of motion only, i.e., it is constant along the particle world-line. The phase function $\Phi(x)$ is presented by the integral
\begin{equation}
\Phi(x) = \frac{\nu}{\sqrt2} \sqrt{\Pi^2_{(0)}{-}\Pi^2_{\bot}} \int{\frac{dx}{\Pi_{(1)}(x)}}
\label{s18}
\end{equation}
with $\Pi_{(1)}(x)$ given by (\ref{OF16}). In more details we obtain
\begin{equation}
\Phi(x) = - \frac{\nu}{\sqrt2}(\tau-\tau_{*}) \sqrt{1+ \frac{\Pi^2_{||}}{m^2}} \ \,.
\label{s19}
\end{equation}

\subsubsection{Transversal motion}

Let the constant of motion $K_3$ be equal to zero. This means that $\Pi_{(3)}(x){=}0$ for arbitrary  $x$, i.e., the longitudinal (with respect to the direction of axis of the universe rotation) component of particle momentum is vanishing, the particle moves in the plane $x^1Ox^2$. There are some simplifications in the formulas for the spin in this case:
\begin{eqnarray}
S_{(0)}(x) &=& S_{*} \frac{\Pi_{\bot}}{m}  \cos{\Phi(x)}\,,
\label{aS15}\\
S_{(1)}(x) &=&
S_{*}\left[\frac{{\cal E}}{m} \cos{\Psi}\cos{\Phi} {-} \sin{\Psi}\sin{\Phi} \right],
\label{aS16}\\
S_{(2)}(x) &=&
S_{*}\left[\frac{{\cal E}}{m} \sin{\Psi}\cos{\Phi} {+} \cos{\Psi}\sin{\Phi} \right]\,,
\label{aS17}
\end{eqnarray}
where we introduced the new auxiliary constant
\begin{equation}
S_{*}= \sqrt{{\cal S}^2-S^2_{(3)}}
\,.
\label{as18}
\end{equation}

\subsubsection{Longitudinal motion}

Let us consider the case, when the transversal integral of motion vanishes, i.e., $\Pi_{\bot}{=}0$. This means that for arbitrary $x$ the physical components of transversal momentum vanish, $\Pi_{(1)}(x){=}\Pi_{(2)}(x)=0$. Since $p_1{=}\Pi_{(1)}(x){=}0$ the particle does not move in the $Ox$ direction. As for the direction $Oy$, we see that $p_2{=}{\cal E} e^{\nu x} \neq 0$. The particle motion in the direction $Oz$ is uniform, $p_3{=}\Pi_{||}$. Since $p_1=0$ we have no possibility to link the variable $x$ and the parameter along the particle world-line $\tau$, and now the procedure of the integration of the master equations for the spin evolution differs  from the procedure described above. For the motion with $\Pi_{\bot}{=}0$ the orthogonality condition $p_k S^k{=}0$ reduces to ${\cal E} S_{(0)}{=} \Pi_{||} S_{(3)}$ with constant values  $S_{(0)}$ and $S_{(3)}$. For simplicity, we put these physical components of the spin four-vector equal to zero, $S_{(0)}{=}S_{(3)}{=}0$, and then the evolution of $S_{(1)}$, $S_{(2)}$ components are guided by the simple system of equations:
\begin{equation}
\frac{dS_{(1)}}{d\tau} = - \omega_{*} \ S_{(2)} \,, \quad \frac{dS_{(2)}}{d\tau} = \omega_{*} \ S_{(1)} \,,
\label{kds7}
\end{equation}
where the constant $\omega_{*}$ is defined as
\begin{equation}
\omega_{*} \equiv  \frac{\nu}{\sqrt2 m} \left(\alpha {+} {\cal E} \right)  \,.
\label{kkds7}
\end{equation}
The corresponding solutions
\begin{eqnarray}
S_{(1)}(\tau) &=& S_{(1)}(0) \cos{\omega_{*} \tau} - S_{(2)}(0)\sin{\omega_{*}\tau}  \,,
\nonumber\\
S_{(2)}(\tau) &=& S_{(2)}(0) \cos{\omega_{*} \tau} + S_{(1)}(0) \sin{\omega_{*}\tau}  \,,
\label{ds7}
\end{eqnarray}
describe the spin precession with the constant angular velocity $\omega_{*}$, and with the amplitude
\begin{equation}
S_{\bot} {=}\sqrt{S^2_{(1)}(0){+}S^2_{(2)}(0)} = {\cal S} \,,
\label{vds7}
\end{equation}
which coincides with the normalization constant ${\cal S}$.

\subsection{Dynamics of massless particle and polarization rotation}

 When $m{=}0$ and thus $k^i k_i =0$, we have to use the coefficient $\alpha_2$ instead $\alpha/m$, which we introduced in the Section \ref{spd in GASu} for convenience.
The master equations, which we have to solve, are of the form:
\begin{eqnarray}
&&\!\!\!\!\!\!\!\!\!\!\!\!\!\!
\frac{dk_i}{d\tau} {-} \frac{1}{2} k^j k^k \partial_i g_{jk} =
\frac{1}{2} \nu \alpha_2 e^{\nu x} \left(\delta_i^1 k^2 {-}\delta_i^2 k^1 \right) {+} \mu k_i \,,
\label{Eq_k}\\
&&\!\!\!\!\!\!\!\!\!\!\!\!\!\!
\frac{d\xi_i}{d\tau} - \frac{1}{2} \xi^j k^l \left[\delta_l^1
g^{\prime}_{ij}+ \delta_i^1 g^{\prime}_{lj} - \delta_j^1
g^{\prime}_{il}\right] =
\nonumber\\&&\qquad\qquad
= \frac{1}{2} \alpha_2 \nu e^{\nu x}
\left(\delta_i^1 \xi^2 -\delta_i^2 \xi^1 \right) + \tilde{\mu} k_i \,.
\label{Eq_xi}
\end{eqnarray}
When $\alpha_2=0$ and $\mu \neq 0$, the equation (\ref{1d1}) yields
\begin{equation}
\frac{d}{d\tau} {\cal E} =   \mu {\cal E}   \ \Rightarrow \ {\cal E}(\tau) =  {\cal E}(0) e^{\mu \tau} \,.
 \label{k1d1}
\end{equation}
Clearly, the physically motivated model has to be characterized by vanishing value of this parameter, $\mu {=}0$.

\subsubsection{Polarization in case of arbitrary direction of motion}

When the massless particle has non-vanishing longitudinal and transversal components of the momentum, we can write the results of integration of the basic equations as follows. First, we use the gauge transformation (\ref{jmm2}) with ${\cal R}{=}{-}\int d\tau \tilde{\mu}$ to eliminate the term $\tilde{\mu} k_i$ in the equation (\ref{Eq_xi}) (see (\ref{j1mm}) for details). Then we take the solutions     (\ref{OFqq}) and (\ref{s15})--(\ref{s17}) and rewrite them using the notations $\Pi_{(i)} \to \pi_{(i)}$, $S_{(i)} \to s_{(i)}$, $\Psi \to \psi$, $\Phi \to \phi$, respectively, for the physical components of the momentum vector  $k_i$ and of the polarization $\xi_i$, as well as, the following simplifications:
\begin{equation}
\sqrt{\pi^2_{(0)}{-}\pi^2_{\bot}} = \pi_{||} \,, \quad {\cal A} =  \frac{1}{\pi_{\bot}\pi_{||}} \,.
\label{004}
\end{equation}
The corresponding solutions are
\begin{eqnarray}
s_{(0)} &=& \frac{1}{\pi_{||}} \left[ s_{(3)} \ \pi_{(0)} + \pi_{\bot}  \cos{\phi(x)} \right]\,,
\label{as15}\\
s_{(1)} &=& \frac{1}{\pi_{||}} \left[s_{(3)} \pi_{(1)} + \right.
\nonumber\\&&\!\!\!\!\!\!\!\!\! + \left.
\pi_{(0)}\cos{\psi} \cos{\phi(x)} {-} \pi_{||} \sin{\psi} \sin{\phi(x)} \right],
\label{as16}\\
s_{(2)} &=&  \frac{1}{\pi_{||}} \left[s_{(3)}\pi_{(2)} + \right.
\nonumber\\&& \!\!\!\!\!\!\!\!\!+\left.
\pi_{(0)} \sin{\psi}\cos{\phi(x)} {+} \pi_{||} \cos{\psi}\sin{\phi(x)} \right].
\label{as17}
\end{eqnarray}
The phase function $\phi(x)$ is now given by
\begin{equation}
\phi(x) = \frac{\nu}{\sqrt2}  \int{\frac{dx}{\pi_{(1)}(x)}}= - \frac{\nu}{\sqrt2}(\tau-\tau_{*})\,.
\label{phi_ml}
\end{equation}
As usual, the component  $s_{(3)}$ remains constant. There are two special cases in the polarization dynamics.

\subsubsection{Longitudinal particle motion}

When $\pi_{\bot}{=}0$ and thus $\pi_{(1)}{=} \pi_{(2)}{=}0$, $\pi_{(0)}{=} \pi_{||}$, the formulas (\ref{as16}) - (\ref{as17}) for the transversal components of spin simplify as follows:
\begin{equation}
s_{(1)} {=} \cos{[\psi {+}\phi]} \,, \quad s_{(2)}{=}  \sin{[\psi {+}\phi]} \,.
\label{vas}
\end{equation}
Clearly, $s^2_{(1)}{+}s^2_{(2)} {=}1$. Also, we see from (\ref{as15}) that $s_{(0)} {=} s_{(3)}$. We can use again the gauge transformation
(\ref{jmm2}) with the following specifications
$$
{\cal R}{=}{-}\frac{\xi_{(0)}}{\pi_{(0)}} \,, \quad \tilde{\mu} = \mu =0 \,, \quad \tilde{\xi}_{(1)} = \xi_{(1)} \,, \quad \tilde{\xi}_{(2)} = \xi_{(2)} \,,
$$
\begin{equation}
\tilde{\xi}_{(0)} = \xi_{(0)} + {\cal R} \pi_{(0)} \,, \quad \tilde{\xi}_{(3)} = \xi_{(3)} + {\cal R} \pi_{||}  \,,
\label{jmm2n}
\end{equation}
thus avoiding the non-physical components of the polarization four-vector, $\tilde{\xi}_{(0)} {=} \tilde{\xi}_{(3)}{=}0$.
Equivalently, the polarization precession can be described by the formulas (\ref{ds7}), in which the quantity
\begin{equation}
\omega_{*} \equiv  \frac{\nu}{\sqrt2} \left(\alpha_2 {+} {\cal E} \right)
\label{vas2}
\end{equation}
plays the role of the precession frequency; for the case of massless particle it becomes the frequency of the polarization rotation.

\subsubsection{Transversal particle motion}

When $\pi_{||}{=}0$, we can not use, formally speaking, the formulas (\ref{as15})--(\ref{as17}), since the vanishing term $\pi_{||}$ is in the denominator.
In order to discuss the corresponding solutions we can use the following tactics. First, one can check directly that the formulas
\begin{eqnarray}
s_{(0)} &=& \pi_{(0)}\sqrt{1-s_{(3)}^2}\, \left(\frac{\nu \tau}{\sqrt2}\right)\,,
\label{vas3}\\
s_{(1)} &=& \sqrt{1-s_{(3)}^2} \left[ \pi_{(1)}\left(\frac{\nu \tau}{\sqrt2}\right) - \frac{\pi_{(2)}}{\pi_{(0)}}\right]\,,
\label{vas4}\\
s_{(2)} &=& \sqrt{1-s_{(3)}^2} \left[ \pi_{(2)}\left(\frac{\nu \tau}{\sqrt2}\right) + \frac{\pi_{(1)}}{\pi_{(0)}} \right]\,,
\label{vas5}
\end{eqnarray}
give the exact solutions for the case $\pi_{||}{=}0$ and $s_{(3)}{=}$const. Then we can find that the gauge transformation (\ref{jmm2}) with the following specifications
\begin{equation}
{\cal R} {=}  \sqrt{1{-}s_{(3)}^2} \left(\frac{\nu \tau}{\sqrt2}\right) \,,
\quad \tilde{\mu} {=} {-}\frac{d{\cal R}}{d\tau}  \,,
\label{n11}
\end{equation}
gives the solutions, which do not depend on the proper time $\tau$, and are free from the non-physical linear growth of the components of the polarization four-vector:
\begin{eqnarray}
\tilde{s}_{(0)} &=& 0 \,,
\label{mmvas3}\\
\tilde{s}_{(1)} &=& - \frac{\pi_{(2)}}{\pi_{(0)}} \sqrt{1-s_{(3)}^2}\,,
\label{mmvas4}\\
\tilde{s}_{(2)} &=&  \frac{\pi_{(1)}}{\pi_{(0)}} \sqrt{1-s_{(3)}^2} \,.
\label{mmvas5}
\end{eqnarray}
Clearly, the orthogonality/normalization relationships
\begin{eqnarray}
&\displaystyle
\pi_{(0)}s_{(0)}  = \pi_{(1)}s_{(1)}{+} \pi_{(2)}s_{(2)} \,,
&\nonumber\\
&\displaystyle
1= {-} s^2_{(0)} +  s^2_{(1)} +  s^2_{(2)} + s^2_{(3)}
\,,&
\label{vas6}
\end{eqnarray}
are satisfied identically.

\section{Second application: G\"odel-type universe supported by a pure aether with one independent Jacobson's constant}

\subsection{Exact solution to the field equations for the model of pure aether}

Let us consider the G\"odel-type universe without matter substratum, i.e., let us suggest that a pure aether is the source of the universe rotation. As it will be shown below, such possibility exists only if there are some constraints for the coupling constants $C_1$, $C_2$, $C_3$, $C_4$. Let us analyze this case in more details.

\subsubsection{Ansatz about the structure of the unit vector field}

Now we assume, that the velocity four-vector has two non-vanishing components
\begin{eqnarray}
&\displaystyle
U^i = \sqrt{1+v^2} \ \delta^i_0 + v \ \delta^i_1   \,,
&\label{2G1}\\
&\displaystyle
U_k = \sqrt{1+v^2}  \ \delta_k^0 -v \  \delta_k^1 + \sqrt{1+v^2} e^{\nu x} \  \delta_k^2 \,.
\label{2G11}&
\end{eqnarray}
In these terms the quantity $v$ is the constant, which describes the $U^1$ component of the unit vector field. Respectively, the square root $\sqrt{1+v^2}$ is the $U^0$ component of the velocity four-vector, and the parameter
$u \equiv \frac{v}{\sqrt{1+v^2}}$ describes the three-dimensional velocity of the aether. Thus, the aether moves in the direction $Ox$, the parameter $v$ is unlimited, $-\infty < v <\infty$, and the parameter $u$ is restricted by $u^2<1$.
The normalization condition is satisfied for arbitrary constants $v$ and/or $u$.

The covariant derivative $\nabla_iU_k$ contains, formally speaking, all four elements of decomposition (\ref{act3})
\begin{equation}
DU_k =v \sqrt{1+v^2}\,\nu  e^{\nu x}\delta^2_k \,, \quad \Theta =v\nu \,,
\label{2G2}
\end{equation}
\begin{widetext}
$$
\omega_{ik} = \frac{\nu}{2}  (1{+}v^2)^{3/2} e^{\nu x}\left(
\begin{array}{cccc}
 0 & 0 & - u   & 0 \\
 0 & 0 &  1  & 0 \\
 u  & -1 & 0 & 0 \\
 0 & 0 & 0 & 0
\end{array}
\right) \,, \qquad
\sigma_{ij}{=} \frac{\nu v\left(1{+}v^2\right)}{3} \left(
\begin{array}{cccc}
u^2  & {-} u   & {-}\frac{1}{2}   e^{\nu  x} u^2  & 0 \\
 {-} u   &  1    & \frac{1}{2} e^{\nu  x} u  & 0 \\
 {-}\frac{1}{2} e^{\nu  x} u^2  & \frac{1}{2} e^{\nu  x} u   & {-} e^{2 \nu  x} \left(1{+}u^2\right)
   & 0 \\
 0 & 0 & 0 & 1{-}u^2
\end{array}
\right).
$$
\end{widetext}
The square of the vorticity tensor remains constant
\begin{equation}
\omega^{pq} \omega_{pq} = (1+v^2)^2\nu^2 \,,
\label{2G4}
\end{equation}
and the angular velocity four-vector of the aether flow is again directed along the axis $0x^3$:
\begin{equation}
\omega^{*j} \equiv \omega^{*jn}U_n =- \frac{\nu(1+v^2)}{\sqrt2}
\delta^j_3  \,.
\label{2G5}
\end{equation}
It is interesting to mention that the angular velocity of the rotation of the aether flow is again constant, but its value differs  from $\nu$ by the factor $(1{+}v^2)$.
This means that the geometric parameter $\nu$ is not now responsible for the aether rotation, and since $v$ is unlimited, the aether rotation rate can exceed significantly the rate of the universe rotation.

\subsubsection{Reduced equations for the unit dynamic vector field}

For the aether motion of the discussed type the quantities $I^j$ and ${\cal J}^{aj}$ take the form
\begin{equation}
I^j = -C_4 \nu^2 v \sqrt{1+v^2} \left[v \delta^j_0 + \sqrt{1+v^2} \delta^j_1 \right] \,,
\label{2G6}
\end{equation}

\begin{widetext}
$$
{\cal J}^{aj}
= \nu \sqrt{v^2+1} \left(
\begin{array}{cccc}
 \left(2 {C_4} \left(v^2+1\right)-{C_2}\right)  u &
   ({C_1}-{C_3})   & \left(-2
   {C_4} (v^2{+}1)+{C_1}+2 {C_2}+{C_3}\right)u e^{-\nu  x}   & 0
   \\
 \left(2 {C_4} v^2-{C_1}+{C_3}\right)   &
   -{C_2} u   & -  \left(2 {C_4}
   v^2-{C_1}+{C_3}\right)e^{-\nu  x}   & 0 \\
 ({C_1}+2 {C_2}+{C_3})u e^{-\nu  x}  &
   -({C_1}-{C_3}) e^{-\nu  x}  & -2
   ({C_1}+{C_2}+{C_3})u e^{-2 \nu  x}   & 0 \\
 0 & 0 & 0 & -{C_2} u
\end{array}
\right)
 \,.
$$
\end{widetext}

The field equations (\ref{0A1}) can be transformed into two algebraic equations for one unknown quantity $\lambda$
$$
(-C_1+C_3+2C_4v^2)\nu^2 = \lambda\,,
$$
\begin{equation}
 -2C_4 (1+v^2)\nu^2 = \lambda\,.
\label{2G14}
\end{equation}
These equations are compatible, when the parameters $C_1$, $C_2$, $C_3$, $C_4$, $v$ are coupled by the relation
\begin{equation}
2C_4 (1+ 2v^2)= C_1-C_3 \,,
\label{2G14z}
\end{equation}
thus, the quantity $\lambda$ reads
\begin{equation}
 \lambda = (C_3-C_1)\frac{(1+v^2)}{(1+2v^2)}\nu^2  \,.
\label{2G1x4}
\end{equation}
Clearly, searching for $\lambda$ we are facing with two cases: first, when $v$
is expressed in terms of Jacobson's parameters, second, when $v$ is arbitrary; these two versions can be realized as follows.

\vspace{3mm}
\noindent
{\it (i)} $C_4 \neq 0$

\noindent
In this case we obtain
\begin{eqnarray}
&\displaystyle
v^2=\frac{C_1-C_3-2C_4}{4C_4}\,,
&\nonumber\\&\displaystyle
\lambda = -\frac12(C_1-C_3+2C_4)\nu^2\,,
\label{2G15}&
\end{eqnarray}
Jacobson's constants being arbitrary ($C_2$ is hidden).

\vspace{3mm}
\noindent
{\it (ii)} $C_4=0$

\noindent
Now we see that
\begin{equation}
C_1=C_3 \,, \quad C_4 = 0 \ \Rightarrow \lambda=0 \,,
\label{2G16}
\end{equation}
$v$, $C_2$ and $C_3$ are arbitrary.

\noindent
To choose one from  these
two versions we have to study the reduced equations for the gravity field.

\subsubsection{Reduced equations for the gravitational field}

When $C_4 \neq 0$, the gravity field equations are of the form
\begin{eqnarray}
&&\!\!\!\!\!\!\!\!\!
\frac12 \nu^2  = \frac{\nu^2}{16C_4} \left[-C_1^2+7C_3^2+12C_4^2 -4C_3C_4- \right.
\nonumber\\&&
\left. -2C_1(C_2+3C_3-10C_4)+2C_2(C_3+2C_4)\right] \,,
\nonumber\\&&\!\!\!\!\!\!\!\!\!
\frac12 \nu^2  e^{\nu x} = \frac{\nu^2}{16C_4} \left[-C_1^2+7C_3^2+12C_4^2 -4C_3C_4-
\right.
\nonumber\\&&
\left. -2C_1(C_2+3C_3-10C_4)+2C_2(C_3+2C_4)\right] e^{\nu x} \,,
\nonumber\\&&\!\!\!\!\!\!\!\!\!
\frac12 \nu^2  =
\frac{\nu^2}{16C_4} \left[C_1^2+(C_3-2C_4)^2+ \right.
\nonumber\\&&
\left. +2C_1(C_2-C_3+2C_4)-2C_2(C_3+2C_4\right]  \,,
\label{2G33}\\&&\!\!\!\!\!\!\!\!\!
\frac34 \nu^2 e^{2\nu x}  = \frac{\nu^2}{32C_4} \left[-C_1^2+15C_3^2+28C_4^2 -12C_3C_4 - \right.
\nonumber\\&&
\left. -2C_1(C_2+7C_3-22C_4)+2C_2(C_3+2C_4)\right] e^{2\nu x}\,,
\nonumber\\&&\!\!\!\!\!\!\!\!\!
\frac12 \nu^2    =-
\frac{\nu^2}{16C_4} \left[C_1^2+(C_3-2C_4)^2- \right.
\nonumber\\&&
\left. -2C_1(C_2+C_3-2C_4)+2C_2(C_3+2C_4) \right]\,.
\nonumber
\end{eqnarray}
There is one important consequence of this set of equations, namely
\begin{equation}
\frac{\nu^2}{8C_4}(C_1-C_3+2C_4)^2=0 \,.
\label{2G34}
\end{equation}
Clearly, in combination with (\ref{2G14z}) this consequence leads to $v^2=-1$, i.e., the parameter $v$ is imaginary. Thus the case $C_4\ne 0$ should be excluded from consideration.

When $C_4 {=} 0$,  the constants $C_1$ and $C_3$ coincides, $C_1{=}C_3$, and $\lambda{=}0$; in this case there are two independent equations in the set of gravity field equations, which can be written as
\begin{equation}
1 = - v^2 (C_2+4C_3) \,, \quad 1 = v^2 C_2 \,.
\label{n1}
\end{equation}
Clearly, the parameter $v$ can be expressed through $C_2$ as follows: $v^2= \frac{1}{C_2}$.
In other words, the solution of the total set of master equations with given ansatz for the velocity four-vector exists, when
\begin{eqnarray}
&\displaystyle
C_1=C_3=-\frac12 C_2 \,, \quad C_4=0 \,,
&\nonumber\\&\displaystyle
\lambda =0 \,, \quad v= \pm \frac{1}{\sqrt{C_2}} \,.
\label{2G37}&
\end{eqnarray}
The positive coupling constant $C_2$ remains arbitrary in the value, and the constants $C_1$ and $C_3$ are negative.

\subsubsection{The velocities of propagation of scalar, vectorial and tensorial modes in the aether}

The interesting detail of this model concerns propagation velocities of scalar, vectorial and tensorial modes (or, for short, of modes spin-0, spin-1, spin-2, respectively). As it was mentioned in \cite{J7}, the squares of the three-dimensional velocities can be represented via Jacobson's constants by the following formulas:
\begin{eqnarray}
&\displaystyle
a^2_{(0)} = \frac{(C_1{+}C_2{+}C_3)(2{-}C_1{-}C_4)}{(C_1{+}C_4)(1{-}C_1{-}C_3)(2{+}C_1{+}C_3{+}3C_2)} \,,
\label{0sp}
&\\
&\displaystyle
a^2_{(1)} = \frac{2C_1-C_1^2+C_3^2}{2(C_1+C_4)(1-C_1-C_3)} \,,
\label{1sp}
&\\
&\displaystyle
a^2_{(2)} = \frac{1}{1-C_1-C_3} \,.
\label{2sp}&
\end{eqnarray}
In the first application, there was no restrictions for these velocities, since when the universe evolution is guided by aether and quasi-dust substratum, the Jacobson's constants were considered to be arbitrary parameters. In the universe supported by the pure aether the mentioned mode velocities have the form
\begin{equation}
a^2_{(0)} = 0 \,, \quad
a^2_{(1)} = a^2_{(2)} {=} \frac{1}{1{+}C_2} {=} \frac{v^2}{1{+}v^2} \equiv u^2
\,.
\label{2sp1}
\end{equation}
In other words, the scalar mode is stopped (suppressed), and the vectorial and tensorial modes  propagate with equal velocities, which coincide with the three-velocity of the aether motion.

\subsection{Spin-particle dynamics in the G\"odel-type universe supported by the pure aether}

In the previous section we have studied the model, in which the aether flow was characterized by the vorticity tensor $\omega_{ik}$ only, and this elegant model admits exact explicit solutions to the equations of particle dynamics and spin (polarization) evolution. Now we discuss the second background model for which all the irreducible parts of the covariant derivative of the velocity four-vector are non-vanishing. As a consequence, the solutions to the equations of particle dynamics and spin (polarization) evolution are much more sophisticated, and we do not intend to discuss their general solutions for arbitrary set of coupling parameters $\alpha_1, ...\alpha_7$, etc. Since our goal is to study the effects of spin precession and of polarization rotation, induced by the coupling to the non-uniformly moving aether, we simplify our task as follows. Let five coefficients from seven in the formula (\ref{OmegaForm}) be equal to zero
\begin{equation}
\alpha_3 = \alpha_4 = \alpha_5 = \alpha_6 = \alpha_7 =0 \,, \label{000}
\end{equation}
and other two be linked by
\begin{equation}
2\alpha_1=\alpha_2=\frac{\alpha}{m}\,.
\end{equation}
Then the tensor $\Omega_{ik}$  contains the contributions from the vorticity tensor $\omega_{ik}$ (as in the previous case) and additional contribution from the acceleration four-vector $DU_i$
\begin{equation}
\Omega_{ik}=\frac{\alpha}{m} \left[\frac12 \left(U_{i}DU_k-U_{k}DU_i\right) + \omega_{ik}\right]\,.
\label{001}
\end{equation}
Again the antisymmetric tensor $\Omega_{ik}$ is characterized by only one non-vanishing component
\begin{equation}
\Omega_{12}=\frac{\alpha \nu}{2m}\ \sqrt{1+v^2}\ e^{\nu x} \,,
\label{002}
\end{equation}
so the particle momentum and spin four-vectors evolution are governed by the equations
\begin{eqnarray}
&&\frac{Dp_i}{D\tau}=\frac{\alpha\nu}{2m} \sqrt{1+v^2} \ e^{\nu x} \left(\delta_i^1 p^2 -\delta_i^2 p^1 \right)\,,\\
&&\frac{DS_i}{D\tau}=\frac{\alpha\nu}{2m} \sqrt{1+v^2} \ e^{\nu x} \left(\delta_i^1 S^2 -\delta_i^2 S^1 \right)\,.
\end{eqnarray}
The right hand side of these equations differ from (\ref{OF4}) and (\ref{s1}) by the factor $\sqrt{1+v^2}$ only. This means that for the description of the particle dynamics and spin evolution we can use the solutions (\ref{OFqq}), (\ref{OF16}), and (\ref{s15})--(\ref{s19}), in which the coefficient $\alpha$ is replaced with $\alpha \sqrt{1+v^2}$.
Similar results are obtained for massless particles with polarization; in that case we have to replace the coupling constant $\alpha_2$ with  $\alpha_2 \sqrt{1+v^2}$.

\section{Discussion}

The unit dynamic vector field, which is associated with the velocity of the aether flow, plays a twofold role in the modeling of the universe evolution.
The first aspect of this role concerns the creation of a specific surrounding space-time, which inherits the properties of the aether flow. Generally, the aether flow is non-uniform and inhomogeneous, i.e., it is characterized by the acceleration, shear, vorticity and expansion. In the first application we focused on the case, when only  vorticity of the aether flow is non-vanishing, and the aether angular velocity is constant. In the second application we assumed that the acceleration, shear and expansion are also non-vanishing.
The master equations for the gravity field (see (\ref{0Ein1})--(\ref{E12})) admit the exact solutions of the G\"odel-type with the metric (\ref{G1}), which can be divided into two classes. The solutions of the first class assume the presence of a matter as a counterpart of the vector field, the second source of the gravitational field. For instance,
the classical solution obtained by G\"odel in \cite{Godel0} is associated with two sources: the dust matter with constant energy density and the cosmological constant. Consistency of the corresponding Einstein equations is provided by the parameters fine-tuning: the dust energy density and the cosmological constant are specifically expressed in terms of the universe rotation parameter $\nu$.
The solutions of the second class are obtained with the assumption that the vector field is the unique source of the gravity field, but the vector field possesses specific internal properties.
The solutions of the first class are illustrated in Section III, where a new exact solution to the master equations for the gravity field is found, which describes the space-time of the G\"odel-type supported by the dynamic aether and stabilized by a matter substratum with anisotropic pressure. The solutions of the second class are discussed in Section IY; they are admissible, when
the scalar modes in the aether are suppressed and the vectorial and tensorial modes propagate with equal velocities, which coincide with the three-velocity of the aether motion as a whole.

The second aspect of the role of the aether flow is related with various marker-effects, which occur in the spatially inhomogeneous non-uniformly moving aether. Clearly, speaking about an aether we imagine some medium, and the aether flow is associated in our mind with its velocity field, which can be laminar, can contain vortex, can be characterized by shear, acceleration and expansion, as the fluid flows, which we see in the life. Marker-systems such as test particles with spin or polarization can signalize to observers, that the aether flow is rotating, by two channels. The first channel is indicated as a geodesic precession; this type of spin precession or polarization rotation is caused by the universe rotation as a whole; it is a global channel of information.    Also, the vortex in the aether flow can interact directly with test marker systems, thus displaying the modus operandi of forces, which act on the test relativistic particle in the aether flow.
There are two typical markers which can bring the global and local information about the aether rotation:  massive test particles with spin and massless test particles with polarization (e.g., photons). In the presented work, we have modeled the corresponding forces for both relativistic massive and massless particles, and solved the equations of the particle dynamics, spin precession and polarization rotation. The main results are the following.

If the coupling constant $\alpha$ in the Lorentz-type forces (\ref{OF3}), (\ref{OF3s}) vanishes, there is no direct influence of the aether on the particle, thus the test relativistic massive particle with spin or massless particle with polarization move along geodesic lines thus monitoring the structure of the gravity field created by the aether with vorticity. In this submodel the exact solutions for the components of the particle momentum (see (\ref{OF10})--(\ref{OF16})) can be interpreted in terms of Magnus effect. Indeed, we can consider the shifts in the $p_1$ and $p_2$ components (appeared only if $\nu \neq 0$) as known hydrodynamic effects in the rotating fluid flow occurring in the plane orthogonal to the fluid rotation axis. We also found that the character of the spin evolution can be indicated as a precession. The angular frequency of the spin precession (and of the polarization rotation for the massless particle) is predetermined by the parameter $\nu$ and depends essentially on the character of particle motion. This can be illustrated, for instance, by the formulas (\ref{s15})--(\ref{s19}) with $\alpha{=}0$; this type of precession can be indicated as geodesic precession, but we understand that the parameter $\nu$  is now originated from the vorticity tensor characterizing the aether flow.

If the coupling constant $\alpha$ is non-vanishing, i.e., there exists a direct influence of the aether flow on the particle motion and spin (polarization) evolution, the situation becomes more sophisticated. We deal now with particle rotation around the axis of the aether vortex; this rotation looks like the rotation of an electrically charged particle moving in the magnetic field (see (\ref{OF15}), (\ref{OF16})). The spin precession in this case can be indicated as hybrid precession. Indeed, it is a composition of the geodesic precession and rotation caused by the Lorentz-type force produced by the vortex in the aether flow; e.g., the hybrid frequency of the spin precession is given by (\ref{kkds7}).

When we deal with massless particles (e.g., photons) with polarization, one can speak about polarization rotation instead of spin precession. Generally, the formulas describing these two processes are similar, nevertheless, there is a specific case, when the massless particle moves in the plane orthogonal to the axis pointing the direction of the rotation of the aether flow. As it was shown in the Section III (compare (\ref{as15})--(\ref{as17}) with (\ref{mmvas3})--(\ref{mmvas5})), we faced with a specific situation, which appears also in ultrarelativistic systems of spinning particles \cite{Der}.

When the aether flow is characterized by acceleration, shear, expansion in addition to the vorticity, the particle motion and the spin evolution become much more sophisticated. We discussed in this paper only one exact solution of this type in order to illustrate one important idea. In Section IV we have shown that, when the aether velocity has the non-vanishing component in the direction orthogonal to the axis of the aether flow rotation, the frequency of the spin precession can be much bigger than the frequency of the aether flow rotation.

We consider the presented theoretical results as a basis for new constraints on the coupling constants $C_1$, $C_2$, $C_3$, $C_4$ and $\alpha$; however, this task is beyond of the scope of this work, and we hope to study this problem in future.

\vspace{4mm}

\acknowledgments{The work was supported by Russian Science Foundation (Project No. 16-12-10401), and, partially, by the Program of Competitive Growth
of Kazan Federal University (Project 0615/06.15.02302.034).}

\end{document}